\documentclass[aps,preprint,superscriptaddress,prl,showpacs,floatfix]{revtex4}
\usepackage{epsfig}
\usepackage{graphicx}
\usepackage{amssymb}
\usepackage{amsmath}

\usepackage{epstopdf}

\begin{document}

\title{Dynamics of uniaxial hard ellipsoids.}
\author{Cristiano~De Michele}
\affiliation{Dipartimento di Fisica and INFM-CRS Soft,
         Universit\`a di Roma {\em La Sapienza}, P.le A. Moro 2, 00185 Roma, Italy}
\author{Rolf~Schilling}
\affiliation{Johannes-Gutenberg-Universitat Mainz, D-55099 Mainz,
Germany}
\author{Francesco~Sciortino} 
\affiliation{Dipartimento di Fisica and INFM-CRS Soft,
         Universit\`a di Roma {\em La Sapienza}, P.le A. Moro 2, 00185 Roma, Italy}
\date{\today}
\begin{abstract}
We study the dynamics of monodisperse hard ellipsoids  via a new event-driven molecular dynamics algorithm as
a function of volume fraction $\phi$ and aspect ratio $X_0$.   We evaluate the
translational $D_{trans}$ and the rotational $D_{rot}$ diffusion coefficient  and the 
associated isodiffusivity lines in the $\phi-X_0$ plane.  We observe a decoupling of the translational and 
rotational dynamics which generates an almost perpendicular crossing of  the $D_{trans}$ and $D_{rot}$ isodiffusivity lines.   While the self intermediate scattering function exhibits stretched relaxation, i.e. glassy dynamics, only for large $\phi$ and $X_0 \approx 1$,  the second order orientational correlator $C_2(t)$ shows
stretching only for large and small $X_0$ values. We discuss these findings in the context of a possible pre-nematic order driven glass transition.

\end{abstract}
\pacs{64.70.Pf,61.20.Ja,61.25.Em,61.20.Lc}
\maketitle
Particles interacting with only excluded volume interaction may exhibit a rich
phase diagram, despite the absence of any attraction.  Spherical objects, in equilibrium, present only a fluid and a crystal phase, while simple non-spherical hard-core particles can form  either crystalline or liquid crystalline ordered phases\cite{allenReview}, as first shown analytically  by Onsager\cite{Onsager} for rod-like particles.  Successive works have established detailed phase diagrams for several  hard-body shapes\cite{Parsons,Lee2,allenHEPhaseDiag,MargoEvans} and have clarified the role of the entropy in the transition between different phases. Less detailed information are available concerning dynamic properties
of hard-core bodies and their kinetically arrested states. In the case of the hard-sphere system, dynamics
slows down significantly on increasing packing fraction $\phi$,
and, when crystallization is avoided (mostly due to intrinsic sample
polydispersity), a dynamic arrested state (a glass) with extremely long life time
can be generated.    The slowing down of the dynamics is well described by
mode coupling theory (MCT)\cite{GoetzeMCT}.  On going from spheres to non-spherical particles, 
non-trivial phenomena arise, due to the interplay between translational and rotational degrees of freedom.  The slowing down of the dynamics can indeed
appear either in both  translational and  rotational properties or in just one of  the two. 

\begin{figure}[tbh]
\vskip 1cm
\includegraphics[width=.48\textwidth]{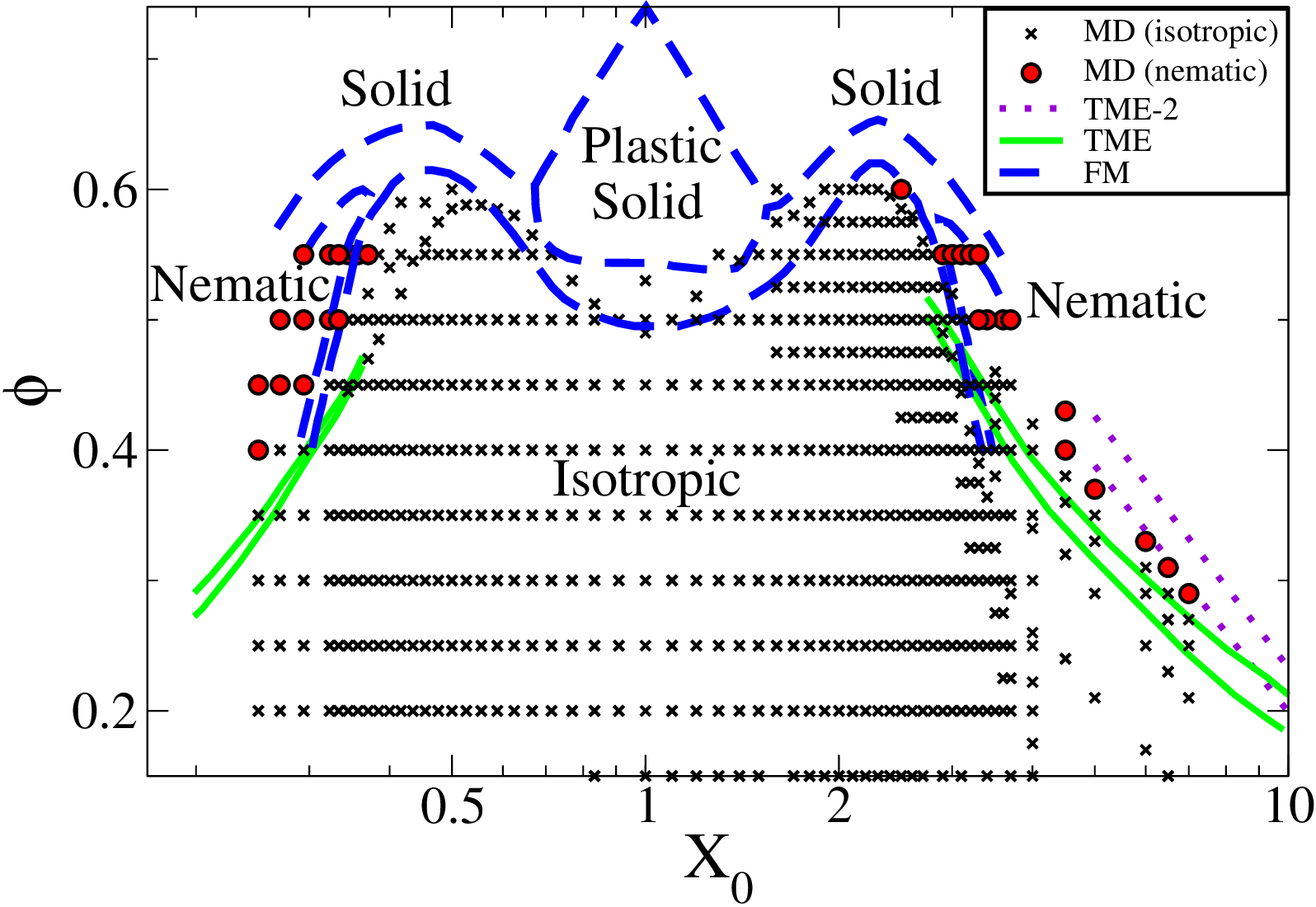}
\caption{Grid of state points simulated (crosses and red-filled circles) 
and relevant boundary lines of coexistence regions.
Long-dashed curves are coexistence curves of all first order phase transitions in the phase diagram of HE
evaluated by Frenkel and Mulder (FM)\cite{FrenkelPhaseDiagMolPhys}. Solid lines are 
coexistence curves for the I-N transition of oblate and prolate ellipsoids, 
obtained analytically by Tijpto-Margo and Evans \cite{MargoEvans} (TME). Dotted lines (TME-2)    
are coexistence curves of prolate ellipsoids for the I-N transition, taken from \cite{MargoEvans}.
} 
\label{Fig:grid}
\end{figure}

Hard ellipsoids (HE)
of revolution \cite{allenReview,singh01} are one 
of the most prominent systems composed by  hard body anisotropic particles.
HE are  characterized by the aspect ratio $X_0=a/b$ (where $a$ is the length of the revolution 
axis, $b$ of the two others) and by the packing fraction $\phi=\pi X_0 b^3 N / 6V$,  where $N$ is the number of particles and $V$ the volume.  The equilibrium phase diagram, evaluated numerically two decades ago 
\cite{FrenkelPhaseDiagMolPhys}, shows an isotropic fluid phase (I) and several ordered phases
(plastic solid, solid, nematic N). The coexistence lines show a swallow-like dependence with a minimum at the spherical limit $X_0=1$ and  a maximum at $X_0\approx 0.5$ and $X_0 \approx 2$ (cf. Figure \ref{Fig:grid}).
Application to HE\cite{LetzSchilLatz} of the molecular MCT (MMCT)\cite{A1ng,A2ng} predicts also  a swallow-like glass transition line. 
In addition, the theory suggests that for $X_0 \lessapprox 0.5$ and $X_0 \gtrapprox 2$, the glass transition is driven by a precursor of nematic order, resulting in an orientational glass where the translational density fluctuations are quasi-ergodic, except for very small wave vectors $q$.
Within MCT, dynamic slowing down associated  to a glass transition is driven by the amplitude of the static correlations.
Since the approach of the nematic transition line is accompanied by an increase of the nematic order correlation function at $q=0$, the non-linear feedback mechanism of MCT results in a glass transition, already before macroscopic nematic order occurs \cite{LetzSchilLatz}.  In the arrested state, rotational motions become hindered.

The recently evaluated random close packing line 
\cite{DonevScience}  for HE also exhibits a swallow-like shape. 
Although Kerr effect measurements in the isotropic phase of liquid crystals have given some evidence for the
existence of two types of glass transitions \cite{A3ng} (related to nematic phase formation and to cage effect
respectively) almost nothing is known about the glassy dynamics of system forming liquid crystals in 
general and for HE in particular. 

\begin{figure}[tbh]
\vskip 1cm
\includegraphics[width=.47\textwidth]{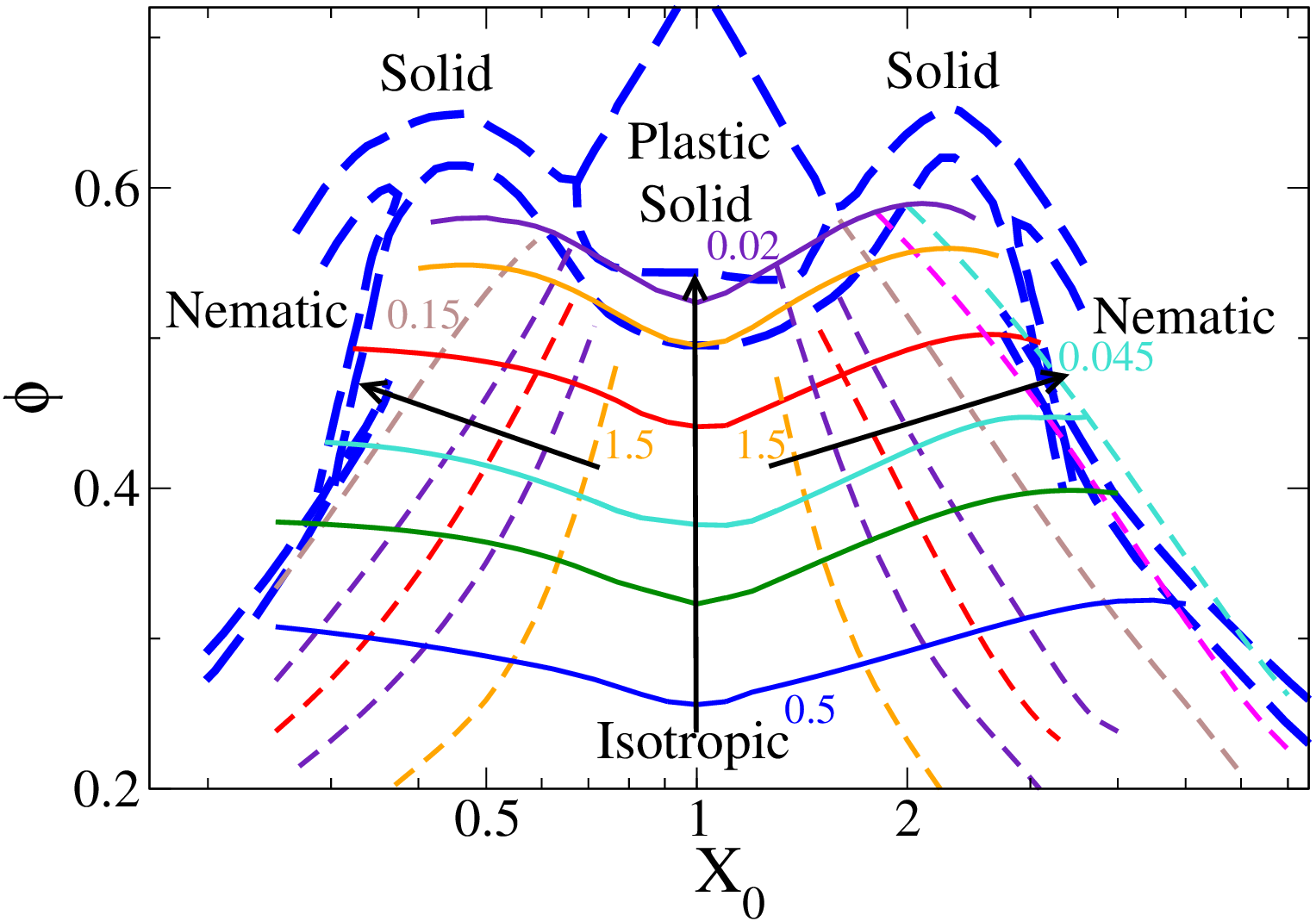}
\caption{Isodiffusivity lines. Solid lines are isodiffusivity lines from translational diffusion coefficients
$D_{trans}$ and dashed lines are isodiffusivities lines from rotational diffusion coefficients $D_{rot}$.
Arrows indicate decreasing diffusivities. Left and right arrows refer to rotational diffusion 
coefficients. Diffusivities along left arrow are: $1.5$, $0.75$, $0.45$, $0.3$, $0.15$.
Diffusivities along right arrow are: $1.5$, $0.75$, $0.45$, $0.3$, $0.15$, $0.075$, $0.045$. 
Central arrow refers to 
translational diffusion coefficients, whose values are: $0.5$, $0.3$, $0.2$, $0.1$, $0.04$, $0.02$.
Thick long-dashed lines are FM and TME coexistence lines from Fig.\ref{Fig:grid}}.
\label{Fig:isod}
\end{figure}

We perform an extended study of the dynamics of monodisperse HE
in a wide window of $\phi$ and $X_0$ values, extending the range of $X_0$
previously studied\cite{AllenFrenkelDyn}. We specifically focus on establishing the
trends leading to dynamic slowing down in both translations and rotations,
by  evaluating  the loci of constant translational and rotational diffusion. These lines, in the limit of vanishing diffusivities, approach the glass-transition lines.
We also study translational and rotational correlation functions, to search for
the onset of slowing down and stretching in the decay of the correlation. 
We perform event-driven (ED)  molecular dynamics simulations, using a new algorithm\cite{DeMicheleScala}, which differently from previous algorithms~\cite{AllenFrenkelDyn,DonevTorqStill1}, relies on evaluations of distance between objects of arbitrary shape.   We simulate a system of $N=512$ ellipsoids at various volumes $V=L^3$ in a cubic box of edge $L$ with periodic boundary conditions.  
We chose the geometric mean of the axis $l=\sqrt[3]{ab^{2}}$ as unit
of distance, the mass $m$ of the particle as unit of mass ($m=1$) and
 $k_BT=1$ (where  $k_{B}$ is the Boltzmann constant and $T$ is the temperature) and hence  the corresponding unit of time is $\sqrt{ml^{2}/k_{B}T}$.
The inertia tensor is chosen as $I_{x}=I_{y}= 2mr^{2}/5$, where $r=\min\{a,b\}$. The value of the $I_{z}$ component 
is irrelevant\cite{allenfrenkelBook}, since  the angular velocity along the symmetry (z-) axis of the HE is conserved. 
We simulate a grid of more than 500 state points at different $X_0$ and $\phi$ as shown in Fig.~\ref{Fig:grid}. To create the starting configuration at a desired $\phi$, we generate a random distribution of ellipsoids at very low $\phi$ and then we progressively  decrease $L$ up to the desired $\phi$. 
We then equilibrate the configuration by propagating the trajectory for times such that 
both angular and translational correlation functions have decayed to zero. 
Finally, we perform a production run at least $30$ times longer than the time needed to equilibrate. For the points close to the I-N transition we check the nematic order by evaluating the largest eigenvalue $S$ of the order tensor ${\bf Q}$~\cite{SpheroCylRec}, whose components are: 
\begin{equation}
Q_{\alpha\beta} = \frac{3}{2}\frac{1}{N}\sum_i \langle({\bf u}_i)_{\alpha} ({\bf u}_i)_{\beta}\rangle - \frac{1}{3} \delta_{\alpha,\beta}
\end{equation}
where $\alpha\beta\in\{x,y,z\}$, and the unit vector $({\bf u}_i(t))_\alpha$ is the component $\alpha$ 
of the orientation (i.e. the symmetry axis) of ellipsoid $i$ at time $t$.
The largest eigenvalue $S$ is non-zero if the system is nematic and $0$
if it is isotropic.  In the following, we choose  the value $S = 0.3$ as criteria to separate isotropic from
nematic states. 
From the grid of simulated state points we build a corresponding grid of
translational  ($D_{trans}$) and diffusional  ($D_{rot}$) coefficients, defined as:
\begin{equation}
D_{trans} = \lim_{t\rightarrow+\infty}  \frac{1}{N} \sum_i \frac{\langle \|{\bf x}_i(t) - {\bf x}_i(0)\|^2  \rangle}{6 t}  
\end{equation}
\begin{equation}
D_{rot} =  \lim_{t\rightarrow+\infty}  \frac{1}{N} \sum_i \frac{\langle \| \Delta\Phi_i \|^2  \rangle}{4 t}  
\end{equation}
where $\Delta\Phi_i = \int_0^t {\bf \omega}_i dt$,
${\bf x}_i$ is position of the center of mass and $ {\bf \omega}_i$  is the angular velocity of ellipsoid $i$.
By proper interpolation, we evaluate the 
isodiffusivity lines,  shown in Fig. \ref{Fig:isod}. 
Results show a striking decoupling of the translational and rotational
dynamics. While the translational isodiffusivity lines 
mimic the swallow-like shape  of the coexistence between the isotropic liquid and the crystalline phases (as well as the MMCT prediction for the glass transition\cite{LetzSchilLatz}), rotational isodiffusivity lines reproduce qualitatively the shape of the I-N coexistence. 
As a consequence of the the swallow-like shape, at large fixed $\phi$, $D_{trans}$   increases by increasing the particle's  anisotropy, reaching its maximum at  $X_0\approx 0.5$ and $X_0\approx 2$. 
Further increase of the anisotropy results in a decrease of $D_{trans}$.  For all $X_0$, an increase of $\phi$ at constant $X_0$ leads to  a significant suppression of $D_{trans}$, demonstrating that $D_{trans}$ is controlled
by packing.
The iso-rotational lines are instead mostly controlled  by $X_0$, showing a progressive slowing down of the rotational dynamics independently from the translational behavior. This suggests that on moving along a path of
constant $D_{trans}$, it is possible to progressively decrease the rotational dynamics, up to the point where rotational diffusion arrest and all rotational motions become hindered.
Unfortunately, in the case of monodisperse HE, a
nematic transition intervenes well before this point is reached. It is thus stimulating to think about the possibility of designing a system of hard particles in which the nematic transition is inhibited by a proper
choice of the disorder in the particle's shape/elongations. 
We note that the slowing down of the rotational dynamics is consistent with MMCT predictions of a nematic glass for large $X_0$ HE\cite{LetzSchilLatz}, in which  orientational degrees of freedom start to freeze approaching the isotropic-nematic transition line, while translational degrees of freedom mostly remain ergodic.
To support the possibility that the slowing down of the dynamics 
on approaching the nematic phase originates from a close-by glass transition,
we evaluate  the self part of the intermediate scattering function $F_{self}$ 
\begin{equation}
F_{self}(q,t) = \frac{1}{N} \langle \sum_j e^{i{\bf q}\cdot({\bf x}_j(t) - {\bf x}_j(0))} \rangle 
\end{equation}
and  the second order orientational correlation function
 $C_2(t)$   defined as \cite{AllenFrenkelDyn} 
$C_2(t) = \langle P_2(\cos\theta(t))\rangle $,
where $P_2(x) = (3 x^2 - 1) / 2$ and $\theta(t)$ is the angle between the symmetry axis at time $t$ and at time $0$.
The $C_2(t)$ rotational isochrones  
are found to be very similar to rotational isodiffusivity lines.  

\begin{figure}[tbh]
\includegraphics[width=.5\textwidth]{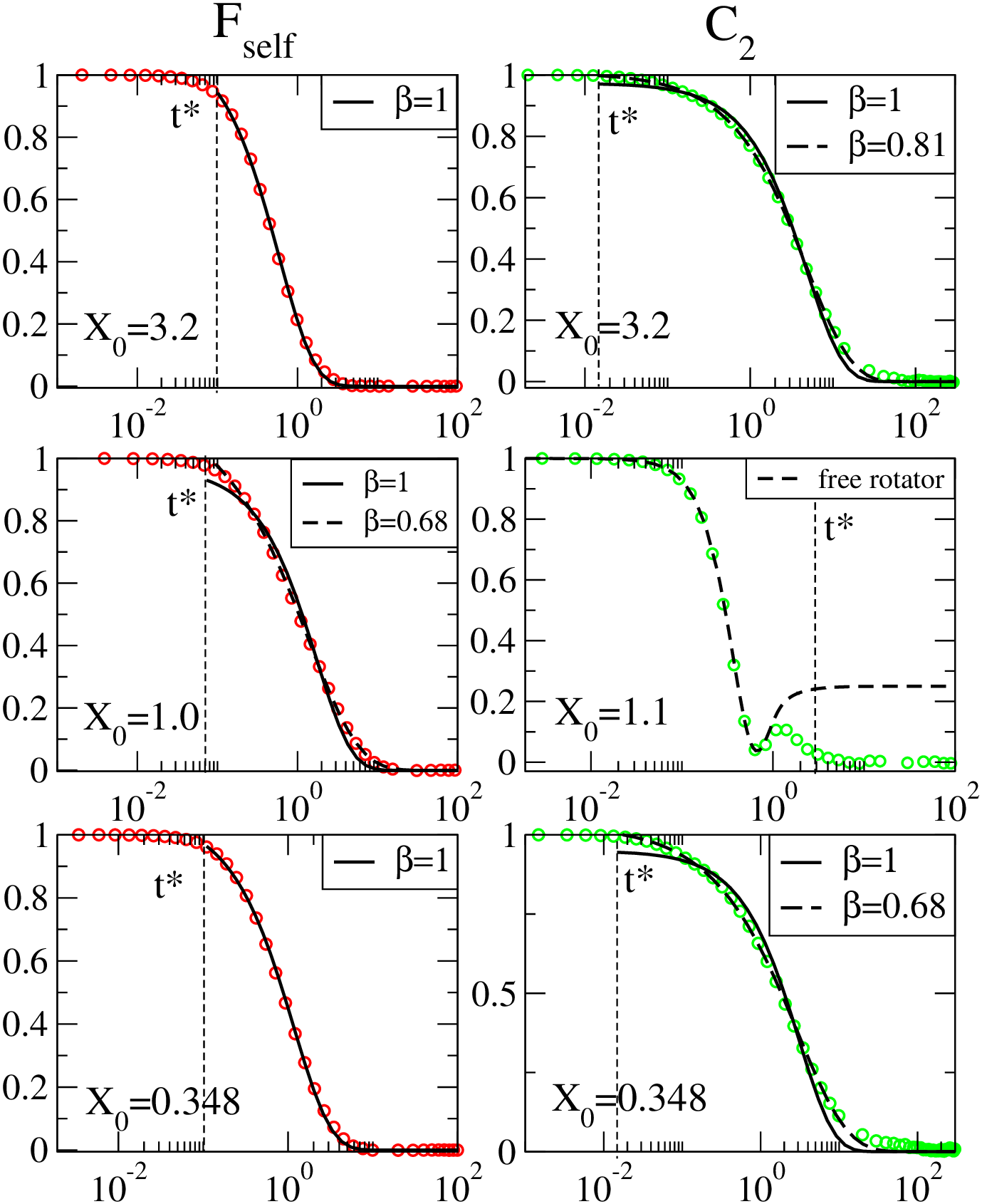}
\caption{Shape of $F_{self}$ and $C_2$ at $\phi=0.50$ for different 
$X_0$. Symbols are data from MD simulations.
Solid lines are fits to exponential functions, while long-dashed lines are fits to
stretched exponentials ($\beta$ is the stretching parameter).    
$t^*$ is the time at which correlation functions $\phi_{vv}$ and $\phi_{\omega\omega}$, for $F_{self}$ and $C_2$
respectively, reach $1/e$ of their initial values.
Top:  Prolate ellipsoids with $X_0=3.2$, $C_2$ shows a significant stretching while $F_{self}$ decays exponentially.
Center: $X_0=1.0$ for $F_{self}$ and $X_0=1.1$ for $C_2$, the dashed line is 
the theoretical decay of a free rotator $C_2^f$  
($C_2^f(t) = 1-\frac{3}{2}\frac{t}{\tau_f}\exp[-t^2/\tau_{f}^2]\tilde{\Phi}(t/\tau_f)$, where 
$\tau_{f}^2=1/\phi_{\omega\omega}(0)$ and $\tilde{\Phi}(t)=\int_0^t \exp[x^2] dx$).
Bottom: Oblate ellipsoids with $X_0=0.348$. }
\label{Fig:corrshape}
\end{figure}

These two correlation functions never show a clear two-step relaxation decay in the entire 
studied  region, even  where the isotropic phase is metastable, since the system can not be
significantly over-compressed. 
As for the well known hard-sphere case, the amount of over-compressing 
achievable in a monodisperse system is rather limited.  This notwithstanding, a comparison of
the rotational and translational correlation functions reveals that the onset of dynamic slowing down and glassy 
dynamics can be detected by the appearance of stretching. 
Fig. \ref{Fig:corrshape} contrasts the shape of  $F_{self}$, evaluated at $q=q_{max}$, where $q_{max}$ is 
the $q$ corresponding to the first maximum of the center-of-mass static structure factor, and  $C_2(t)$ 
at $\phi=0.50$ for different $X_0$ values with best-fit based on
an exponential  ($ \sim \exp[-t/\tau]$) and a stretched exponential ($ \sim \exp[-(t/\tau)^\beta]$) decay. As a criteria to avoid including in the fit the short-time ballistic contribution, we limit the time-window to 
times larger than $t^*$,  defined for  $F_{self}$ and $C_2$ 
as the time at which the autocorrelation 
function of center-of-mass velocity ${\bf v}$  (${\phi}_{vv}(t) \equiv \frac{1}{N}\sum_i \langle {\bf v}_i(t){\bf v}_i(0) \rangle$ ) and of angular velocity respectively
(${\phi}_{\omega\omega}(t) \equiv \frac{1}{N}\sum_i  \langle {\bf\omega}_i(t) {\bf\omega}_i(0) \rangle$) reaches $1/e$ of its initial value.
We note that $F_{self}$ shows an exponential behaviour close to the I-N transition ($X_0=3.2$,$\ 0.3448$) 
on the prolate and oblate side,  in agreement
with the fact that translational isodiffusivities lines do not exhibit any peculiar behaviour close to the I-N line. Only when $X_0\approx 1$, 
$F_{self}$ develops a small stretching, consistent with the minimum of the swallow-like curve observed in the 
fluid-crystal line \cite{HardSpheresExp,HardSpheresSim},  in the jamming locus as well as  in the predicted behavior of the glass line for HE\cite{LetzSchilLatz} and for small elongation dumbbells\cite{DumbellChongGoetze,DumbellChongFra}.  Opposite behavior is seen for the case of the 
orientational correlators. $C_2$ shows stretching at large anisotropy, i.e. at small and large $X_0$ values,
but decays within the microscopic time for almost spherical particles.  
In this quasi-spherical limit, the decay is well represented by the decay of  a free rotator\cite{FreeRotator}.  Previous studies of the rotational dynamics of 
HE\cite{AllenFrenkelDyn} did not report  stretching in $C_2$, probably due to the smaller values of $X_0$ previously investigated and to the present increased statistic  which allows us to follow the full decay of the correlation functions.

\begin{figure}[tbh]
\vskip 1cm
\includegraphics[width=.45\textwidth]{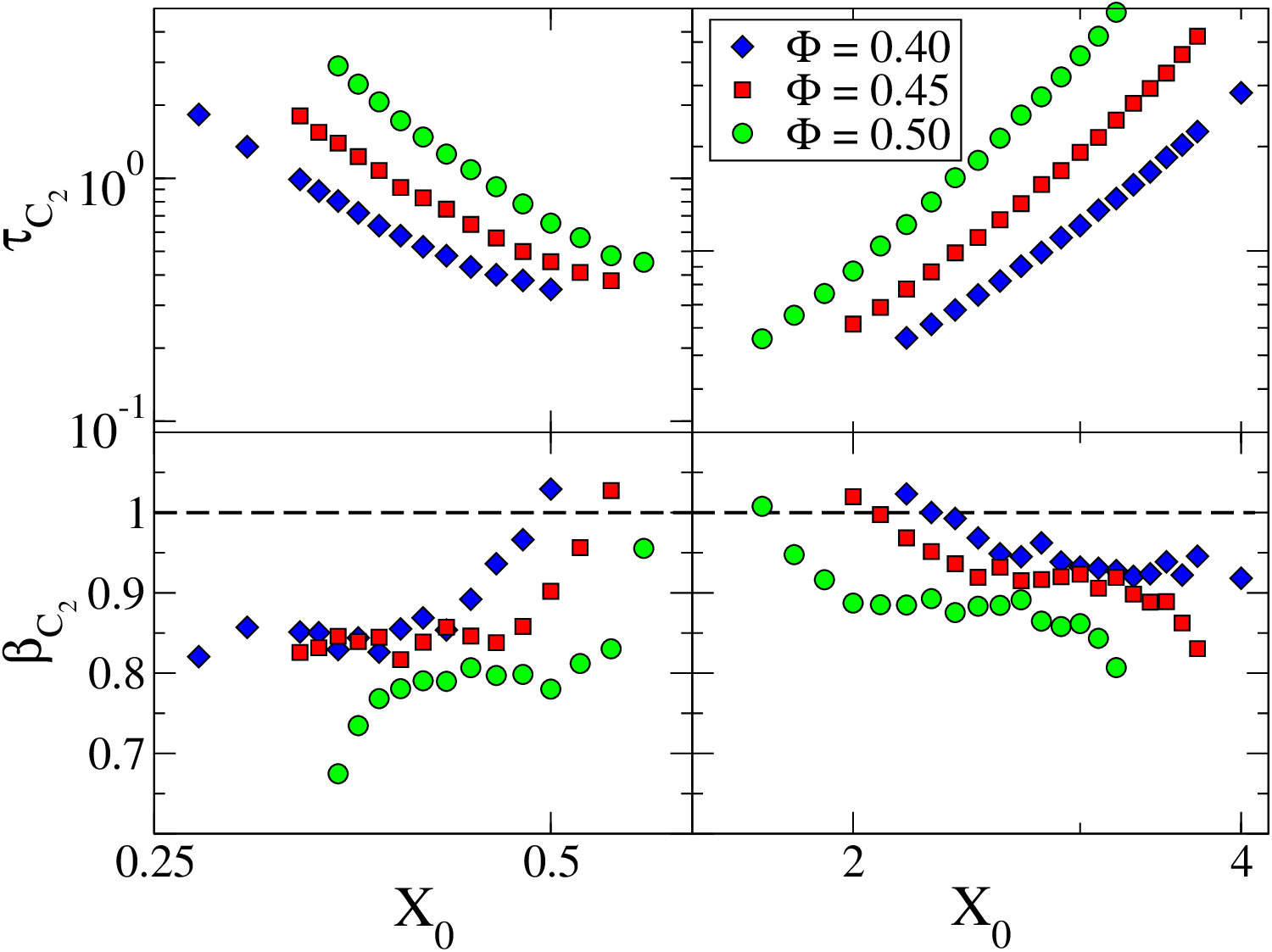}
\caption{$\beta_{C_2}$ and $\tau_{C_2}$ are obtained from fits of $C_2$ to a stretched exponential for $\phi=0.40,0.45$ and $0.50$. Top: $\tau_{C_2}$ as a function 
of $X_0$. Bottom: $\beta_{C_2}$ as a function of $X_0$. The time window used 
for the fits is chosen in such a way to exclude the microscopic short times ballistic
relaxation (see text for details). For $0.588 < X_0 < 1.7$ the orientational relaxation is 
exponential.}
\label{Fig:betavsX0}
\end{figure}

Fig. \ref{Fig:corrshape} clearly shows that  $C_2$ becomes stretched approaching the I-N transition while $F_{self}$  remains exponential on approaching the  transition.
To quantify the amount of stretching in $C_2$ we show  in Fig. \ref{Fig:betavsX0}
the $X_0$ dependence of $\tau$ and $\beta$ for three different values of $\phi$.
In all cases, slowing down of the characteristic time and stretching increases progressively on approaching the  I-N transition. It is interesting to observe that the amount of stretching appears to be more pronounced in the case of oblate HE compared to prolate ones.
A similar (slight) asymmetry between oblate and prolate HE can be observed in the  lines reported in Figure $2$.

In summary, we have shown that clear precursors of dynamic slowing down and stretching  can be observed in  the region of the  phase diagram where a (meta)stable isotropic phase can be studied. Despite the monodisperse character of the present system prevents the possibility of observing a clear glassy dynamics, 
our data  suggest that a slowing down in the orientation degree of freedom ---  driven by the elongation of the particles --- is in action. The main effect of this shape-dependent slowing down is 
a decoupling of the translational and rotational dynamics which generates an almost perpendicular crossing of the $D_{trans}$ and $D_{rot}$ isodiffusivity lines.
This behavior is in accordance with  MMCT predictions, suggesting two  glass transition mechanisms, related respectively to cage effect (active for $0.5 \lessapprox X_0 \lessapprox 2$) and to pre-nematic order ($X_0 \lessapprox 0.5$, $X_0 \gtrapprox 2$) \cite{LetzSchilLatz}. 
It remains to be answered if it is possible to find a suitable model, for example 
polydisperse in size and elongation, for which nematization can be sufficiently destabilized,  in analogy to the destabilization of crystallization induced by polydispersity in hard-spheres.  
We acknowledge support from MIUR-PRIN. 
We also thank A. Scala for suggesting  code optimization and 
taking part to the very early stage of this project.
\bibliography{nemglass.bib}
\end{document}